\documentclass[journal]{IEEEtran}
\usepackage{blindtext, graphicx}

\usepackage{balance}  
\usepackage{fixltx2e}
\usepackage{algorithmicx}
\usepackage{algorithm}
\usepackage[noend]{algpseudocode}
\usepackage{url}
\let\labelindent\relax
\usepackage{enumitem}
\usepackage{multirow}
\usepackage{graphicx}
\usepackage{balance}  
\usepackage{url}
\usepackage[acronym,toc,shortcuts]{glossaries}
\usepackage{epsfig}
\usepackage{amssymb}
\usepackage{amsthm}
\usepackage{multirow}
\usepackage{caption}
\usepackage{subcaption}
\usepackage{siunitx}
\usepackage{soul}
\usepackage{algorithm}
\usepackage{algpseudocode}
\usepackage{csquotes}
\usepackage{bbm}
\usepackage{dsfont}
\usepackage{xcolor}
\usepackage{enumitem}
\usepackage{adjustbox}
\usepackage{array}
\usepackage{multirow}
\usepackage{fancyhdr}
\usepackage[T1]{fontenc}
\usepackage{balance}

\usepackage[acronym,toc,shortcuts]{glossaries}

\graphicspath{{figures/}}

\newacronym{3GPP}{3GPP}{The 3rd Generation Partnership Project }
\newacronym{5G}{5G}{Fifth Generation}
\newacronym{6G}{6G}{Sixth Generation}
\newacronym{AAA}{AAA}{Authentication, Authorization and Accounting}
\newacronym{AES}{AES}{Advanced Encryption System}
\newacronym{AI}{AI}{Artificial Intelligence}
\newacronym{AP}{AP}{Access Point}
\newacronym{API}{API}{Application Programming Interface}
\newacronym{APN}{APN}{Access Point Name}
\newacronym{AR}{AR}{Augmented Reality}
\newacronym{BS}{BS}{Base Station}
\newacronym{BER}{BER}{Bit Error Rate}
\newacronym{BSSID}{BSSID}{Basic Service Set Identification}
\newacronym{CAT}{CAT}{Capacity-Aware TOPSIS}
\newacronym{CEP}{CEP}{Complex Event Processing}
\newacronym{CELL-ID}{CELL-ID}{cell identification ID}
\newacronym{CGI}{CGI}{Cell Global Identification}
\newacronym{CLSM}{CLSM}{Closed loop spatial multiplexing}
\newacronym{CQI}{CQI}{Channel Quality Indicator}
\newacronym{CN}{CN}{core network}
\newacronym{CNN}{CNN}{Convolutional Neural Networks}
\newacronym{CL}{CL}{Closed-Loop}
\newacronym{CoMP}{CoMP}{coordinated multi-point}
\newacronym{CS}{CS}{central scheduler}
\newacronym{CSI}{CSI}{Channel Status Information}
\newacronym{CPU}{CPU}{Central Processing Unit}
\newacronym{CU}{CU}{Central Unit}
\newacronym{eNB}{eNB}{evolved Node-B}
\newacronym{DL}{DL}{Downlink}
\newacronym{DES}{DES}{Data Encryption Standard}
\newacronym{DMM}{DMM}{Distributed Mobility Management}
\newacronym{DoS}{DoS}{Denial of Service}
\newacronym{DTLS}{DTLS}{Datagram Transport Layer Security}
\newacronym{DU}{DU}{Distributed Unit}
\newacronym{EC}{EC}{Edge Computing}
\newacronym{ECA}{ECA}{Event-Condition-Action}
\newacronym{ECC}{ECC}{Elliptic Curve Cryptography}
\newacronym{eNodeB}{eNodeB}{evolved Node-B}
\newacronym{ECDF}{ECDF}{Empirical Cumulative Distribution Function}
\newacronym{E-RAB}{E-RAB}{E-UTRAN Radio Access Bearer}
\newacronym{ETSI}{ETSI}{European Telecommunications Standards Institute}
\newacronym{FDD}{FDD}{Frequency Division Duplexing }
\newacronym{FEM}{FEM}{Flow Extraction Manager}
\newacronym{GGSN}{GGSN}{Gateway GPRS Support Node}
\newacronym{GPRS}{GPRS}{General packet radio service}
\newacronym{GTP}{GTP}{GPRS Tunneling Protocol}
\newacronym{HAPS}{HAPS}{High-Altitude Platform Stations}
\newacronym{HetNet}{HetNet}{heterogeneous network}
\newacronym{HSS}{HSS}{Home Subscriber Station}
\newacronym{HTTP}{HTTP}{Hypertext Transfer Protocol}
\newacronym{HTTPS}{HTTPS}{Hypertext Transfer Protocol Secure}
\newacronym{HDFS}{HDFS}{Hadoop Distributed File System}
\newacronym{HiveQL}{HiveQL}{Hive Query language}
\newacronym{HSPA}{HSPA}{High Speed Packet Access}
\newacronym{IBLER}{IBLER}{Initial Block Error Rate}
\newacronym{ICIC}{ICIC}{inter-cell interference coordination}
\newacronym{ICN}{ICN}{information-centric network}
\newacronym{IEEE}{IEEE}{Institute of Electrical and Electronics Engineers}
\newacronym{IETF}{IETF}{Internet Engineering Task Force}
\newacronym{IMSI}{IMSI}{International Mobile Subscriber Identity}
\newacronym{IMEI}{IMEI}{International Mobile Station Equipment Identity}
\newacronym{IMS}{IMS}{IP Multimedia Subsystem}
\newacronym{ICMP}{ICMP}{Internet Control Message Protocol}
\newacronym{IoT}{IoT}{Internet of Things}
\newacronym{IP}{IP}{Internet Protocol}
\newacronym{IPSec}{IPSec}{Internet Protocol Security}
\newacronym{ISL}{ISL}{Inter-Satellite Link}
\newacronym{ITU}{ITU}{International Telecommunication Union}
\newacronym{IT}{IT}{Information Technology}
\newacronym{GBR}{GBR}{Guaranteed Bit Rate}
\newacronym{GLUE}{GLUE}{General Language Understanding Evaluation}
\newacronym{JSON}{JSON}{JavaScript Object Notation}
\newacronym{KPI}{KPI}{Key Performance Indicator}
\newacronym{LA}{LA}{Location Area}
\newacronym{LAC}{LAC}{location area code}
\newacronym{LMA}{LMA}{Local Mobility Anchor}
\newacronym{LSTM}{LSTM}{ Long-Short Term Memory}
\newacronym{LTE}{LTE}{long term evolution}
\newacronym{MADM}{MADM}{Multiple Attribute Decision Making}
\newacronym{MANO}{MANO}{Management and Orchestration}
\newacronym{MCC}{MCC}{Mobile Country Code}
\newacronym{MEC}{MEC}{Mobile Edge Computing}
\newacronym{MCS}{MCS}{Modulation Coding Scheme}
\newacronym{MCP}{MCP}{Management Control Policy}
\newacronym{MNC}{MNC}{Mobile Network Code}
\newacronym{MIMO}{MIMO}{multiple-input multiple-output}
\newacronym{MAG}{MAG}{Mobile Access Gateway}
\newacronym{MAAR}{MAAR}{Mobility Anchor and Access Router}
\newacronym{ML}{ML}{Machine Learning}
\newacronym{MME}{MME}{Mobility Management Entity}
\newacronym{MN}{MN}{Mobile Node}
\newacronym{MNO}{MNO}{Mobile Network Operator}
\newacronym{MSISDN}{MSISDN}{Mobile Station International Subscriber Directory Number}
\newacronym{NBI}{NBI}{NorthBound Interface}
\newacronym{NE}{NE}{Network Equipment}
\newacronym{NFV}{NFV}{Network Function Virtualization}
\newacronym{NIST}{NIST}{National Institute of Standards and Technology}
\newacronym{NLP}{NLP}{Natural Language Processing}
\newacronym{NLU}{NLU}{Natural Language Understanding}
\newacronym{NOC}{NOC}{Network Operations Center}
\newacronym{NOMA}{NOMA}{Non-Orthogonal Multiple Access}
\newacronym{NoSQL}{NoSQL}{Not Only SQL}
\newacronym{NR}{NR}{New Radio}
\newacronym{NS}{NS}{Network Service}
\newacronym{NTN}{NTN}{Non-Terrestrial Networks}
\newacronym{QoS}{QoS}{quality-of-service}
\newacronym{QoE}{QoE}{quality-of-experience}
\newacronym{OAM}{OAM}{Operation, Administration and Management}
\newacronym{ONF}{ONF}{Open Networking Foundation}
\newacronym{ONOS}{ONOS}{Open Network Operating System}
\newacronym{OS}{OS}{operating system}
\newacronym{OL}{OL}{Open-Loop}
\newacronym{PDN}{PDN}{packet data network}
\newacronym{PF}{PF}{Proportional Fair}
\newacronym{P-GW}{P-GW}{packet gateway}
\newacronym{PDP}{PDP}{Packet Data Protocol}
\newacronym{PHY}{PHY}{physical layer}
\newacronym{PKI}{PKI}{Public Key Infrastructure}
\newacronym{PMIPv6}{PMIPv6}{Proxy Mobile IPv6}
\newacronym{PMI}{PMI}{Precoding Matrix Index}
\newacronym{PRB}{PRB}{Physical Resource Block}
\newacronym{PUSCH}{PUSCH}{Physical Uplink Shared Channel}
\newacronym{REST}{REST}{Representational State Transfer}
\newacronym{QAM}{QAM}{Quadrature amplitude modulation}
\newacronym{QCI}{QCI}{QoS Class Identifier}
\newacronym{O-RAN}{O-RAN}{Open Radio Access Network}
\newacronym{QRSA}{QRSA}{Quantum Resistant Security Algorithm}
\newacronym{rApp}{rApp}{radio App}
\newacronym{RA}{RA}{Routing Area}
\newacronym{RB}{RB}{Resource Block}
\newacronym{RI}{RI}{Rank Indicator}
\newacronym{RU}{RU}{Remote Unit}
\newacronym{RAN}{RAN}{Radio Access Metwork}
\newacronym{RFC}{RFC}{Request for Comment}
\newacronym{RIC}{RIC}{RAN Intelligent Controller}
\newacronym{RRC}{RRC}{Radio Resource Control}
\newacronym{RNC}{RNC}{radio network controller}
\newacronym{RNN}{RNN}{Recurrent Neural Networks}
\newacronym{RSSI}{RSSI}{Received Signal Strength Indicator}
\newacronym{RSRP}{RSRP}{Reference Signal Received Power}
\newacronym{RRM}{RRM}{Radio Resource Management}
\newacronym{RT}{RT}{Real Time}
\newacronym{OTT}{OTT}{over-the-top}
\newacronym{SA}{SA}{Stand Alone}
\newacronym{SAC}{SAC}{service area code}
\newacronym{SCMA}{SCMA}{Sparse Code Multiple Access}
\newacronym{SLA}{SLA}{Service Level Agreement }
\newacronym{SDN}{SDN}{Software Defined Networking}
\newacronym{SDO}{SDO}{Standards Developing Organization}
\newacronym{S-GW}{S-GW}{serving gateway}
\newacronym{SINR}{SINR}{signal-to-interference-plus-noise ratio}
\newacronym{SGSN}{SGSN}{Serving GPRS Support Node}
\newacronym{SCO}{SCO}{Service \& Computation Orchestrator}
\newacronym{SFF}{SFF}{Simple-Feed-Forward}
\newacronym{SSID}{SSID}{Service Set Identification}
\newacronym{SVD}{SVD}{singular value decomposition}
\newacronym{TCP}{TCP}{transport control protocol}
\newacronym{TDD}{TDD}{Time Division Duplexing}
\newacronym{TLS}{TLS}{Transport Layer Security}
\newacronym{TM}{TM}{transmission mode}
\newacronym{TN}{TN}{Terrestrial Network}
\newacronym{TEID}{TEID}{tunnel endpoint identifier}
\newacronym{UAS}{UAS}{Unmanned Aerial Systems}
\newacronym{UDN}{UDN}{Ultra Dense Network}
\newacronym{UMTS}{UMTS}{Universal Mobile Telecommunications Service} 
\newacronym{UE}{UE}{user equipment}
\newacronym{UL}{UL}{Uplink}
\newacronym{UDP}{UDP}{User Datagram Protocol}
\newacronym{V2X}{V2X}{Vehicle-to-everything}
\newacronym{VNF}{VNF}{Virtual Network Function}
\newacronym{WiFi}{WiFi}{Wireless Fidelity}
\newacronym{WLAN}{WLAN}{Wireless Local Area Network}

\begin{document}
%
\title{Probabilistic Forecasting for Network Resource Analysis in Integrated Terrestrial and Non-Terrestrial Networks}

\author{Cristian J. Vaca-Rubio$^{\diamond}$, Vaishnavi Kasuluru$^{\diamond}$,  Engin~Zeydan$^{\diamond}$,  Luis Blanco$^{\diamond}$, Roberto Pereira$^{\diamond}$, \\   Marius Caus$^{\diamond}$  and  Kapal Dev$^{\ast}$  \\

\thanks{This work has been submitted to IEEE for possible publication. Copyright
 may be transferred without notice, after which this version may no longer be
 accessible.}
 $^{\diamond}$Centre Tecnològic de Telecomunicacions de Catalunya (CTTC), Castelldefels, Barcelona, Spain, 08860. \\
 $^{\ast}$CONNECT Centre and Department of Computer Science, Munster Technological University. Ireland. \\
\protect \{cvaca, vkasuluru, ezeydan, lblanco,  rpereira, mcaus\}@cttc.cat, kapal.dev@ieee.org\vspace{-.5cm}}

\maketitle

\begin{abstract}

Efficient resource management is critical for Non-Terrestrial Networks (NTNs) to provide consistent, high-quality service in remote and under-served regions. While traditional single-point prediction methods, such as Long-Short Term Memory (LSTM), have been used in terrestrial networks, they often fall short in NTNs due to the complexity of satellite dynamics, signal latency and coverage variability. Probabilistic forecasting, which quantifies the uncertainties of the predictions, is a robust alternative. In this paper, we evaluate the application of probabilistic forecasting techniques, in particular \ac{SFF}, to NTN resource allocation scenarios. Our results show their effectiveness in predicting bandwidth and capacity requirements in different NTN segments of probabilistic forecasting compared to single-point prediction techniques such as \ac{LSTM}. The results show the potential of \textcolor{black}{probabilistic forecasting models} to provide accurate and reliable predictions and to quantify their uncertainty, making them indispensable for optimizing NTN resource allocation. At the end of the paper, we also present application scenarios and a standardization roadmap for the use of probabilistic forecasting in integrated \textcolor{black}{Terrestrial Network (TN)}-NTN environments.

\end{abstract}

\begin{IEEEkeywords}
Probabilistic forecasting, non-Terrestrial networks, O-RAN, autonomous networks, AI/ML. 
\end{IEEEkeywords}

\IEEEpeerreviewmaketitle


\section{Introduction}

The emergence of \ac{NTN} technologies, potentially including Low/Medium/Geostationary satellites (LEO, MEO, GEO), \ac{HAPS} and \ac{UAS} has revolutionized telecommunications, providing ubiquitous and resilient connectivity \cite{guidotti2024role}. NTNs extend 5G/B5G coverage but face latency, Doppler shifts, and signal attenuation. LEO satellites (500–1500 km) offer low latency \cite{hosseinian2021review} but require efficient resource use. Besides, resource demand varies by time and location, influenced by population density and economic activity. Furthermore, requiring adaptive resource allocation.  At the same time in \glspl{TN}, the emerging concept of \ac{O-RAN} introduces a new paradigm of network design and operation. The result is an open architecture that disaggregates network components, offering the possibility of integrating hardware and software from multiple vendors.  Based on the time-scale, the \ac{O-RAN} architecture can  differentiate between loops that operate at msec latency, within the range of 10msec-1s and above 1s, yielding \ac{RT}, Near-\ac{RT} and Non-\ac{RT} applications \cite{polese2023understanding}.  This feature has the potential to overcome the lack of flexibility of legacy implementations based on monolithic units.

NTN is now part of 3GPP for ubiquitous connectivity in line with the 'anytime, anywhere' paradigm \cite{el2023introduction}. In order to deploy flexible and globally accessible architectures, it is considered necessary to bring the \ac{O-RAN} framework into space \cite{campana2023ran}. However, the \ac{O-RAN} concept has not yet been fully exploited in the context of \ac{NTN}. The impact of satellite propagation delays on O-RAN interfaces and component placement requires further study, which can be either on the ground or onboard. The decision depends mainly on the interface constraints and the trade-off between the communication delay and the computing capabilities of the satellites. The analysis carried out in earlier works, e.g. \cite{Lar19,Cam23,Dau24}, shows that the \ac{RU} and the \ac{DU} cannot be split across the feeder link. This translates into moving the \ac{RU} and the \ac{DU}  to the satellite payload. As for the \ac{CU} and the Near-\ac{RT} \ac{RIC} placement, the reason for implementing these components on the ground is to reduce the energy consumption of the payload, but at the cost of increasing the service latency. This problem is solved by moving the \ac{CU} and the Near-\ac{RT} \ac{RIC} to space, which places a higher demand on on-board processing. Remarkably, the most appropriate placement of \ac{CU} and Near-\ac{RT} \ac{RIC} in \ac{NTN} remains open as it depends on a variety of factors.

Integrating \ac{O-RAN} with \ac{NTN} can offer a framework to apply \ac{AI} to manage the integrated \glspl{RAN}.  For this reason, to unleash the full potential of this framework, it is of utmost importance to understand the network functions that host the training and the AI algorithms as well as the control loops. Integrating both \ac{TN}-\ac{NTN}s can bring several benefits in addition to added complexity, such as increased system overhead and advanced management requirements that require advanced intelligent resource provisioning methods that dynamically adapt to changing network conditions. Probabilistic forecasting can offer a promising solution by predicting resource requirements with quantified uncertainty. By integrating \textcolor{black}{probabilistic forecasting models \cite{haupt2019use}} into the integrated \ac{TN}-\ac{NTN} architecture, operators can efficiently allocate resources such as bandwidth, beam capacity and power in different segments. This paper explores the implementation of probabilistic forecasting in such integrated environments \textcolor{black}{using real-world satellite data} and evaluates its impact on optimizing resource allocation and maintaining quality of service in comparison to single point estimators.

\section{General Architecture and Probabilistic Forecasting}

\subsection{General Architecture}

\begin{figure*}[htp!]
\centering
\includegraphics[width=.6\linewidth]{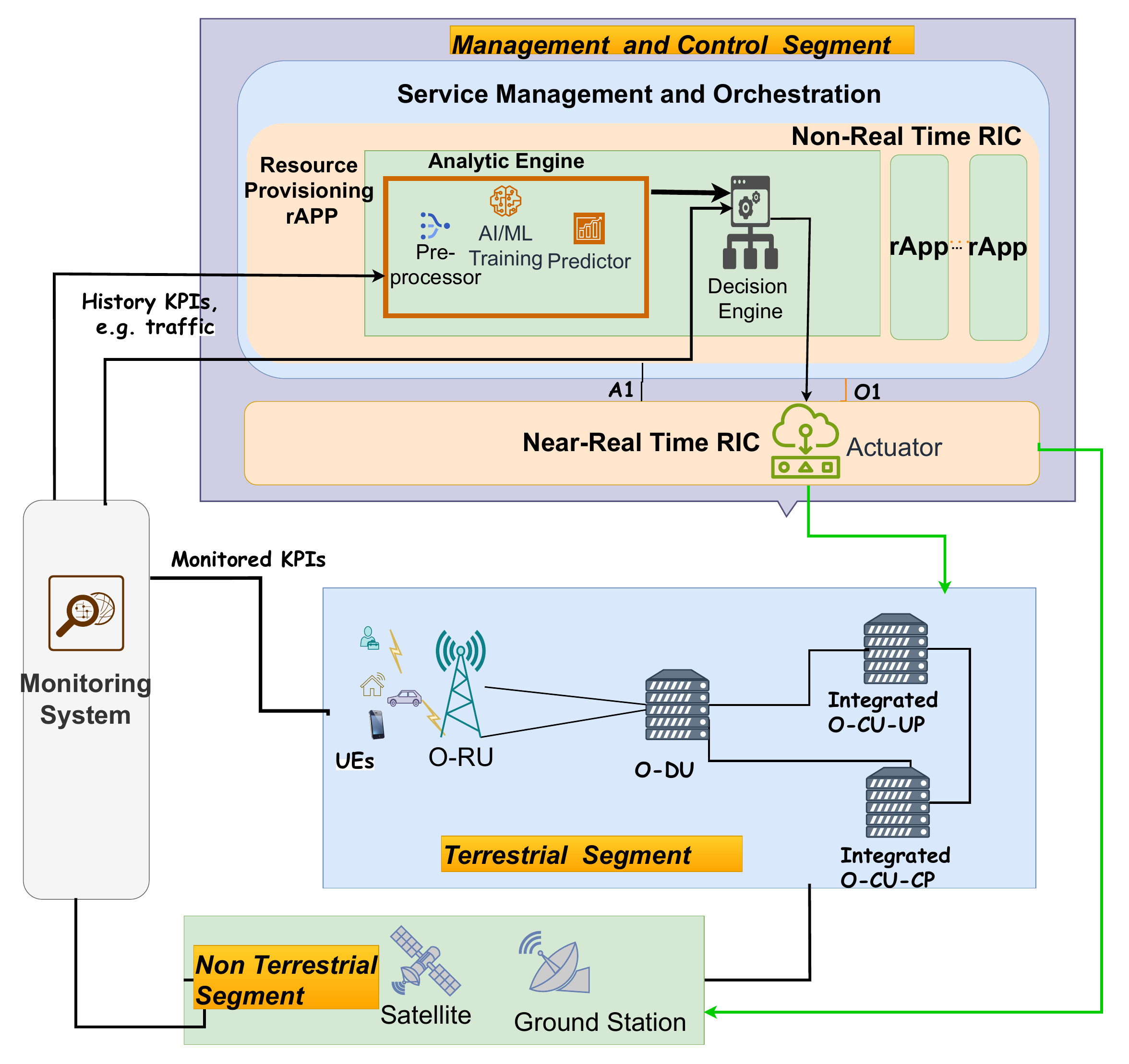}
\caption{Integrated NTN and TN (with Open RAN) architecture with probabilistic forecasting as rApp.}
\label{RAN_New}
\vspace{-0.4cm}
\end{figure*}

Fig. \ref{RAN_New} \textcolor{black}{illustrates the integration of \glspl{NTN} into the \ac{O-RAN} architecture, detailing the interaction between key components across terrestrial, non-terrestrial and management \& control segments.} The \ac{rApp} consists of four sub-modules: the monitoring system, the analytic engine, the decision engine and the actuator. \textcolor{black}{The role of the Non-\ac{RT} \ac{RIC} and Near-\ac{RT} \ac{RIC} in \ac{AI}-driven resource provisioning is also shown, where historical \glspl{KPI} from NTN and terrestrial sources are monitored and processed through the resource provisioning  \ac{rApp} analytic engine \cite{10329915}.} \textcolor{black}{The decision engine and actuator can be  used to dynamically adjust resources for both terrestrial and satellite-based nodes, ensuring proactive resource allocation for \ac{SLA} compliance. The actuator performs optimized resource allocation decisions in terrestrial and \ac{NTN} segments by dynamically adjusting \ac{PRB} allocation, beam steering, ISL routing, power control and frequency configurations to ensure optimal network performance. It follows a hybrid control model that combines centralized decision making in the Non-\ac{RT} \ac{RIC}  and real-time adjustments in the Near-\ac{RT} \ac{RIC} for policy updates.}  In the context of \ac{NTN}, the \ac{rApp} architecture is extended to take into account the unique characteristics and requirements of \ac{NTN} environments, such as satellite movement, beam steering and dynamic coverage areas. Each submodule is adapted to handle these complexities. The monitoring system gathers telemetry data, while the analytic engine predicts resource needs and adjusts allocations dynamically. The decision engine optimizes bandwidth, beam steering, and traffic routing for efficient allocation. The actuator implements these decisions in real time and performs tasks such as beam width adjustment, frequency reallocation and seamless handovers between satellites and ground stations. Probabilistic forecasting improves TN-NTN resources, predicting traffic peaks, handovers, and beam demands. By incorporating probabilistic forecasting, operators can optimize resource utilization, improve network performance and maximize efficiency across terrestrial and non-terrestrial \ac{RAN} components.

\subsection{Probabilistic Forecasting in O-RAN and NTN}
\label{frameworks}

The \ac{O-RAN} framework is a flexible and modular approach to managing network resources in traditional terrestrial networks. Its principles of disaggregation, openness and programmability make it a suitable foundation for the integration of \ac{NTN}s into next generation communication systems. By extending O-RAN to \ac{NTN}s, we can enable efficient resource provisioning in space (satellites), airborne (\ac{HAPS}/\ac{UAS}) and in terrestrial segments, ensuring seamless and scalable operations. We present the adaptation of the O-RAN framework to \ac{NTN}s, focusing on the architecture, resource provisioning components and integration with \ac{NTN}-specific elements such as satellite nodes, \ac{HAPS} and \ac{UAS}.

Probabilistic forecasting models (e.g. DeepAR, Transformer, SFF) that are used for \glspl{TN} previously \cite{kasuluru2023use, kasuluru2024impact} can also be adapted to handle \ac{NTN}-specific data, including satellite motion and handover in LEO orbits, dynamic beam steering, and variable signal quality in \ac{NTN}-based communications. The predictions can also take into account latency, Doppler effects and environmental factors such as weather-related signal attenuation. Unlike deterministic models, probabilistic methods estimate the conditional distribution of future values instead of single-point predictions. In practice, this is typically done by training \textcolor{black}{probabilistic forecasting models} on time-series data while ensuring that they learn the dependencies and underlying processes governing the data. Rather than providing a single prediction per time step, probabilistic methods provide a set of statistical parameters (e.g., mean and variance) that describe the underlying probability distribution at each time step.

These models predict probability distributions, enabling confidence intervals for scenario assessment. A typical use case is to predefine an acceptable range of outcomes (prediction intervals) and determine the probability that the true value will fall within that range. For instance, by predicting the mean and variance of a future value, one can calculate confidence intervals or quality ranges, enabling systems to make risk-aware decisions and adjust dynamically to fluctuations in network performance, signal degradation, or other uncertainties. This probabilistic framework allows for more informed, flexible decision-making compared to deterministic models, particularly in environments where variability and unpredictability play a significant role.
\textcolor{black}{Having established the general architecture and the role of probabilistic forecasting, the following section details the key components and their interplay in the integrated \ac{TN}-\ac{NTN} ecosystem.}

\section{Key Components and Workflow Between TN and NTN Entities}

\subsection{Key Components}

The integrated \ac{TN}-\ac{NTN} frame consists of several key components. The most important of these are as follows:

\subsubsection{Non-RT RIC}

Located in the \ac{NOC} or central cloud, the Non-\ac{RT} \ac{RIC}, processes the collected data for long-term network optimization. It focuses on optimizing resource allocation policies and training AI/ML models based on historical network data and satellite telemetry. The main tasks include (i) training AI/ML models (e.g. DeepAR, Transformer) for the probabilistic forecasting of traffic demand and resource requirements. (ii) Defining policies for resource allocation, \ac{SLA} compliance and multi-tenant prioritization. (iii) Updating control algorithms and sending them to the Near-\ac{RT} \ac{RIC} for real-time implementation. The output is Policies, forecasts, and optimization parameters are forwarded to the Near-\ac{RT} \ac{RIC}.

\subsubsection{Near-RT RIC}
Handles low-latency tasks, such as beam steering, bandwidth allocation, and satellite handovers. It is deployed onboard satellites or at ground stations and incorporates predictive models for managing Doppler shifts and latency.  The Near-\ac{RT} \ac{RIC}, deployed onboard satellites, \ac{HAPS} or at ground gateways, manages real-time tasks that require latency in the millisecond range. Key tasks include (i) dynamic resource allocation for service connections (e.g. bandwidth, beam capacity). (ii) Managing beam steering and satellite handover based on real-time demand predictions. (iii) Ensuring \ac{SLA}s are met by adjusting resource allocation in response to traffic fluctuations. (iv) Monitoring network performance metrics and sending feedback to the Non-\ac{RT} \ac{RIC} for further optimization. 

\subsubsection{gNB and NTN-gNB (Next-Generation NodeB)}

The gNB/\ac{NTN}-gNB serves as a central connection point for user devices to the network via the \ac{NTN} nodes. In the case of transparent payloads, the signals are forwarded without processing; the gNB/\ac{NTN}-gNB in the ground segment takes over all control and processing tasks. For regenerative payloads, the gNB/\ac{NTN}-gNB is partially or fully implemented on board the satellite or \ac{HAPS} and enables signal processing, error correction and routing in real time. \textit{(iv) User Equipment (UE) and Service Provisioning} With Service Links, user devices such as smartphones, IoT sensors or VSAT terminals connect to the \ac{NTN}. The integrated TN and \ac{NTN} framework ensures that resources are dynamically provisioned based on the user’s location and demand and that QoS remains consistent even during satellite handovers or peak traffic times.

\begin{figure*}[htp!]
\centering
\includegraphics[width=.7\linewidth]{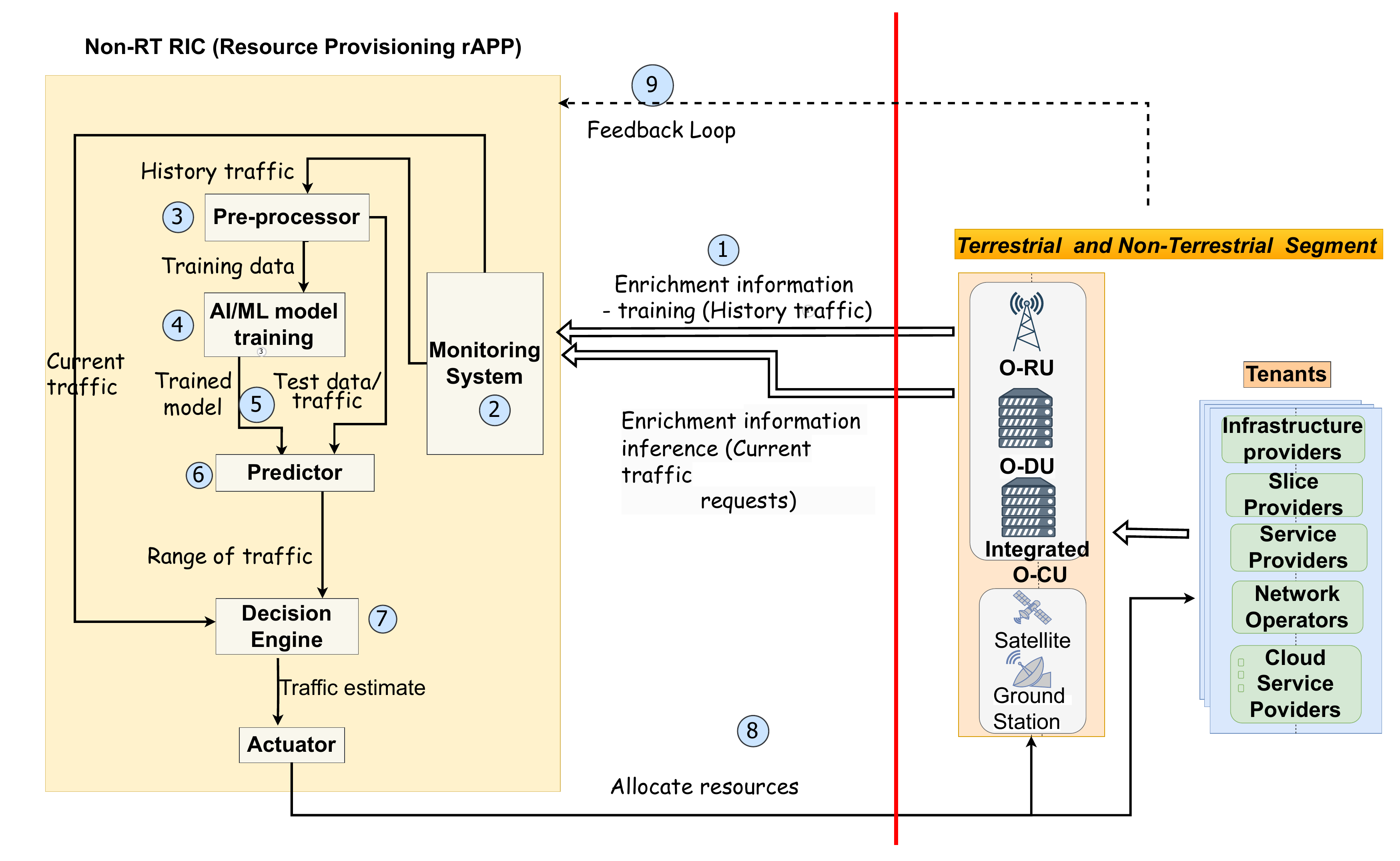}
\caption{Enhanced workflow of integrated \ac{TN}-NTN infrastructure  for Resource Provisioning.}
\label{RAN_Flow}
\vspace{-0.4cm}
\end{figure*}

\subsection{Proposed Workflow for Integrated Networks}

Fig. \ref{RAN_Flow} shows the workflow between different entities in the \ac{O-RAN} framework  integrated with \ac{NTN}, where  tenants/operators demand the network for radio resources, including \ac{PRB}s, bandwidth, and beam allocation in \ac{NTN}s. The proposed workflow leverages Non-\ac{RT} \ac{RIC} for long-term optimization and Near-\ac{RT} \ac{RIC} for real-time operation. Feedback loops have been developed to ensure adaptive and robust performance in various \ac{NTN} scenarios, including multi-tenant and \ac{SLA}-sensitive use cases.

\begin{itemize} 
    \item[\textbf{1.}] Each of the tenants provides \textit{historical resource demand data} (e.g., \ac{PRB}s for \ac{NTN} and TN, and bandwidth or beam demand for nodes) to the monitoring system.
    \item[\textbf{2.}] \textit{Data Collection and Telemetry Gathering (Ground Segment \& \ac{NTN} Nodes)}. The Monitoring System collects telemetry from \ac{TN}-\ac{NTN} nodes (e.g., satellites, HAPS, UAS, O-RU, O-DU), including traffic load/patterns, signal quality, beam utilization, Doppler shifts or  positional information. It also receives historical and real-time data from tenants.  This information is  forwarded to other elements in the rApp for pre-processing.     
    \item[\textbf{3.}] As part of the \textbf{analytic engine}, pre-processing involves aggregating historical data from terrestrial and non-terrestrial sources (e.g., satellites, gateways, ground nodes) and splitting it into training and test data. For \ac{NTN} and TN, this includes aligning data sets with telemetry from dynamic topologies such as LEO constellations, HAPS movements or gNodeB/satellite traffic patterns. 
    \item[\textbf{4.}] The \textit{AI/ML model training} phase uses probabilistic forecasting algorithms, e.g. \ac{SFF}, DeepAR, and Transformer, to handle the complexity of integrated \ac{TN}-\ac{NTN} resource dynamics. The training data from the preprocessor is used to train models that predict future resource requirements, such as beam capacity, bandwidth, and \ac{PRB}s, across terrestrial and non-terrestrial segments.
    \item[\textbf{5.}] The \textit{trained model and the test data} from the preprocessor are passed to the predictor for performance evaluation. In the integrated \ac{TN}-\ac{NTN} context, this includes testing the models for accuracy in predicting resource demands, such as bandwidth requirements for feeder links, beam shifts, \ac{ISL} or gNodeB traffic.
    \item[\textbf{6.}] The \textit{Predictor} returns forecasted estimates of resource demands, including \ac{PRB}s for terrestrial networks and specific integrated \ac{TN}-\ac{NTN} metrics (e.g., beam coverage and bandwidth usage) for a given duration. For example, the prediction can cover a 24-hour window. Forecasts can provide ranges rather than scalar values, offering uncertainty metrics for resource provisioning.      The range of predicted resources includes time-varying values across percentile ranges (99-th to 1-th) based on the \ac{ECDF} of the forecast. This helps to account for uncertainty and variability in integrated \ac{TN}-\ac{NTN}-specific factors such as satellite movement, latency, and coverage overlap.  The Near-\ac{RT} \ac{RIC} can use forecasts to dynamically allocate resources, minimizing both overprovisioning and underprovisioning whereas  the Non-\ac{RT} \ac{RIC} can refine long-term allocation policies based on AI-driven insights, ensuring network scalability and reliability.
    \item[\textbf{7.}] The \textit{Decision engine} receives the estimated ranges of resources with their probabilities and determines the exact number of \ac{PRB}s or other resources (e.g., bandwidth, beam capacity) to allocate to tenants at a given time. In the integrated \ac{TN}-\ac{NTN} context, this includes deciding optimal beam steering or inter-satellite or gNodeB bandwidth distribution based on the predicted traffic and \ac{SLA} requirements.
    \item[\textbf{8.}] The \textit{Actuator} forwards the exact resource allocation values received from the decision engine to the corresponding nodes (e.g., satellites, ground stations, or terrestrial O-DU) via the O1 interface. For integrated \ac{TN}-\ac{NTN}, this includes commands for dynamic beam adjustments, frequency reallocation, or bandwidth assignments. In this setup, the Near-\ac{RT} \ac{RIC} can dynamically allocate resources to tenants based on forecasts and current demand. Tenant-specific allocation can prioritize emergency services, maritime communications or IoT applications based on \ac{SLA} requirements. Traffic Adjustment can redirect underutilized resources from one tenant to another in real time to optimize network performance.
    \item [\textbf{9.}] \textit{Feedback Loop for Continuous Optimization:} Feedback from \ac{TN}-\ac{NTN} nodes and gateways is continuously sent to both the Near-\ac{RT} \ac{RIC} and the Non-\ac{RT} \ac{RIC}. In the Near-\ac{RT} \ac{RIC} feedback, real-time performance metrics and \ac{SLA} compliance reports help refine immediate resource allocation. In Non-\ac{RT} \ac{RIC} feedback, long-term trends and aggregated telemetry data are used to retrain \ac{AI} models and update resource allocation policies. This creates a closed loop for dynamic and adaptive resource provisioning.

\end{itemize}

\section{Experimental Results \& Discussions}
\label{experiments}
We conducted experiments on probabilistic forecasting in TN-NTN, focusing on real-world satellite traffic prediction. Our objective is to demonstrate how these forecasting techniques improve \textit{resource management, network efficiency, and service continuity} across terrestrial and non-terrestrial segments. By leveraging and adapting a Multi-Layer perceptron (MLP) for probabilistic forecasting, that we denote as \ac{SFF} (e.g. we learn the parameters of an assumed Gaussian [$\mu$, $\sigma$] by maximizing the Gaussian log-likelihood) and \ac{LSTM} for deterministic forecasting, we aim to provide a comprehensive an analysis of how forecasting-based decision-making can optimize network operations. We present a case study to illustrate how these forecasting models improve \ac{TN}-\ac{NTN} resource allocation and discuss their implications for potential real-world deployments. \textcolor{black}{Specifically, models use 7-day hourly traffic data to forecast the next 24 hours}.

\textcolor{black}{\subsection{Model selection justification for TN-NTN}}

\textcolor{black}{The SFF model was chosen because it offers a good balance between predictive ability and computational simplicity. In \ac{NTN} environments where hardware constraints and low latency requirements are critical, SFF provides sufficiently accurate predictions while enabling faster inference and easier deployment compared to more complex models such as DeepAR or Transformer \cite{kasuluru2024impact}. These factors are critical in scenarios that require real-time resource allocation decisions under limited onboard processing capabilities.}

\subsection{Case Study: Probabilistic and Deterministic Forecasting for TN-NTN Resource Management}
For this case study, we will focus on the prediction of satellite traffic for the last 24 hours with 1-hour granularity given as historical data the prior 7 days of the week over 6 beam areas.
\subsubsection{Uncertainty quantification in satellite traffic prediction vs deterministic forecasting}
Accurate traffic forecasting is essential for optimizing resource allocation, minimizing service disruptions, and improving network efficiency. Traditional deterministic forecasting methods predict a single-point estimate of future traffic without considering inherent uncertainties, which can lead to critical misallocation of resources. To address this, the \ac{SFF} method is employed, which not only estimates expected traffic but also provides probabilistic confidence intervals, allowing for a more robust and risk-aware decision-making process.
\begin{figure}[t]
\centering
\includegraphics[width=0.9\linewidth]{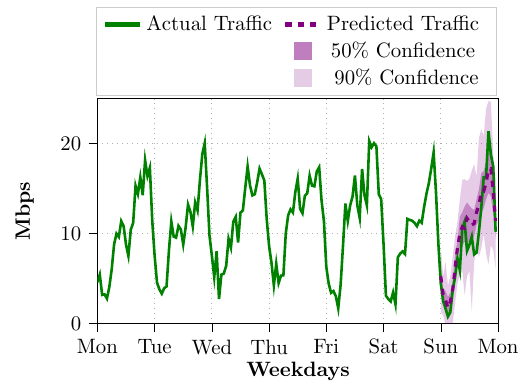}
    \caption{SFF \textcolor{black}{probabilistic} satellite  traffic forecasting.}
\label{fig:sff}
\vspace{-0.4cm}
\end{figure}
Figure \ref{fig:sff} illustrates the predicted satellite traffic using the SFF approach, where the purple-shaded regions represent uncertainty quantification across different probability intervals. The purple line denotes the mean forecasted traffic. The model effectively captures the overall trend in traffic fluctuations, accurately following rapid variations over time. The uncertainty bands (confidence intervals) provide valuable insights into the reliability of predictions. For instance, the 50\% confidence interval (darker shaded region) encompasses most of the data points but does not always capture extreme variations while the 90\% confidence interval (lighter shaded region) correctly encompasses the entire traffic trend, offering higher reliability for risk-averse applications. By incorporating probabilistic forecasting, network operators can make informed decisions based on confidence levels, rather than relying solely on single-point estimates. This flexibility is crucial for adaptive satellite networks. 

\begin{figure}[htp!]
\centering
\includegraphics[width=0.9\linewidth]{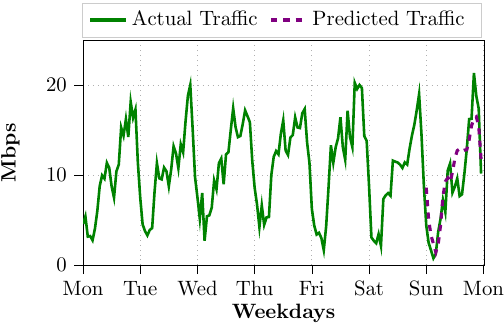}
\caption{LSTM \textcolor{black}{deterministic} satellite traffic forecasting.}
\label{fig:lstm}
\vspace{-0.4cm}
\end{figure}

\begin{figure}[htp!]
\centering
\includegraphics[width=0.95\linewidth]{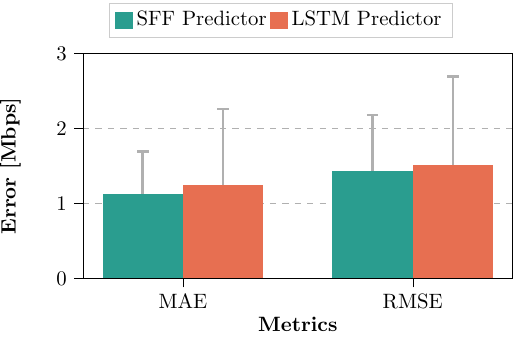}
\caption{Prediction errors for SFF and LSTM.}
\label{fig:errors}
\vspace{-0.4cm}
\end{figure}

In contrast, Figure \ref{fig:lstm} shows the same satellite traffic predictions using a deterministic \ac{LSTM} model. Although the predictions follow a similar overall trend \textcolor{black}{to} the SFF model in \textcolor{black}{the} mean \textcolor{black}{value}, this approach suffers from \textcolor{black}{various} limitations:  1) No \textcolor{black}{quantification of} uncertainty, \textcolor{black}{i.e.} there are no confidence guarantees \textcolor{black}{in connection} with the forecast, 2) single-point predictions can be misleading, as they fail to capture the variability in satellite traffic, leading to potential misallocations of bandwidth and computational resources, and 3) in high-stakes network management scenarios, incorrect predictions without confidence measures can have severe consequences, such as service degradation or inefficient satellite beam switching strategies. To further validate this approach, Figure \ref{fig:errors} quantifies the forecasting accuracy based on Mean Absolute Error (MAE) and Root Mean Square Error (RMSE). The results show that LSTM produces slightly higher errors than SFF. Furthermore, not only SFF accounts for uncertainty predictions, but it performs more accurate predictions in the mean.

\subsubsection{Provisioning performance analysis}
\begin{figure}[t]
\centering
\includegraphics[width=0.95\linewidth]{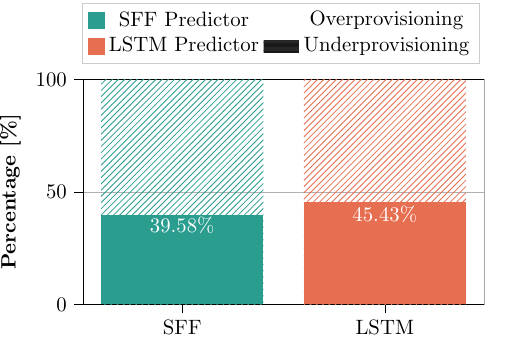}
\caption{Over/under-provisioning for SFF and LSTM.}
\label{fig:provisioning}
\vspace{-0.4cm}
\end{figure}
Figure \ref{fig:provisioning} compares overprovisioning and underprovisioning rates between SFF and LSTM models. As shown in the results, SFF experiences 60.42\% overprovisioning and 39.58\% underprovisioning, whereas LSTM shows 54.17\% overprovisioning and 45.83\% underprovisioning. While LSTM appears to reduce overprovisioning, its higher underprovisioning rate highlights a risk of insufficient resource allocation, potentially leading to service degradation. Probabilistic forecasting with SFF ensures a more balanced approach, mitigating the risk of resource shortages while maintaining efficiency in \ac{TN}-\ac{NTN} environments.

\textcolor{black}{\subsection{Discussions and Practical Considerations}}
\label{discussion}

The results obtained in the case study show the advantages of probabilistic forecasting, especially in predicting the dynamics of satellite traffic. However, for fully integrated \ac{TN}-\ac{NTN} networks, it is critical to extend these forecasting techniques beyond isolated satellite systems to jointly manage resources across terrestrial and non-terrestrial segments. In the following, we discuss how the findings from our experiments can improve \ac{TN}-\ac{NTN} orchestration. \textit{(i) Coordinated allocation of resources:} A major challenge in \ac{TN}-\ac{NTN} is the coordinated allocation of resources across terrestrial and non-terrestrial infrastructures. Traditional methods often consider TN and \ac{NTN} separately, which leads to inefficient utilization. By integrating resource estimation based on probabilistic forecasts, network operators can predict fluctuations in the demand for satellite traffic and use this information to dynamically adjust the allocation of TN resources (e.g. PRBs and bandwidth), to prevent TN congestion events (e.g., urban traffic surges), and to optimise the network. Traffic surges can be preemptively offloaded to \ac{NTN} segments before congestion occurs, or balance the load between satellite backhaul and terrestrial base stations by providing appropriate capacity margins based on uncertainty intervals. \textit{(ii) TN-NTN Load Balancing Application:} If a probabilistic model predicts a surge in satellite traffic, the system can increase TN base station provisioning for ground users who rely on backhaul, pre-allocate additional feeder link bandwidth to avoid congestion in satellite uplinks or activate \ac{NTN} beams in specific regions where terrestrial connectivity is expected to degrade. \textit{(iii) Cross-Segment \ac{SLA} Optimization: } \textcolor{black}{Probabilistic forecasting models} play a crucial role in ensuring \ac{SLA} compliance within \ac{TN}-\ac{NTN} networks by providing uncertainty-aware traffic demand predictions that help operators dynamically manage resources. By offering confidence intervals for traffic variations across terrestrial and non-terrestrial segments, these models enable proactive decision-making to guarantee minimum service levels even under fluctuating conditions. This allows networks to adjust QoS levels in real-time, prioritizing \ac{NTN} resources when terrestrial congestion is anticipated. In \ac{SLA}-sensitive applications, such as maritime or aviation connectivity, probabilistic forecasting minimizes service disruptions by facilitating early resource reallocation between TN and \ac{NTN}, ensuring seamless and reliable coverage for critical communications.

\textcolor{black}{Finally, from the perspective of practical implementation of AI/ML model deployment and computational feasibility, the considered \ac{AI} models, such as \ac{SFF}, can be deployed at different O-RAN levels, with Non-\ac{RT} \ac{RIC} performing long-term training in cloud environments and Near-\ac{RT} \ac{RIC} performing real-time inference on satellite gateways or regenerative payloads for efficient \ac{NTN} resource management. Moreover, given power and bandwidth constraints in \ac{NTN}, strategies such as model quantization to reduce complexity, federated learning for decentralized training, and \ac{AI} acceleration through low-power chips can enable efficient deployment of \ac{AI} models.}

\textcolor{black}{In real NTN deployments, uncertainty quantification is crucial for efficient resource allocation, especially in dynamic LEO satellite environments where coverage and demand fluctuate. Decision makers or AI agents, leveraging probabilistic forecasting, enable adaptive strategies for bandwidth allocation, beam steering, and load balancing, optimizing resource distribution based on confidence intervals. By integrating uncertainty-aware approaches like reinforcement learning and probabilistic scheduling, satellite operators can enhance network resilience and efficiency, ensuring proactive adjustments to traffic demands and channel conditions.}

\section{Standardization Roadmap, Challenges and Future Directions} 
\label{standardization}

\subsection{Standardization Roadmap}

The introduction of probabilistic forecasting methods in \ac{NTN}s requires adaptation to existing and emerging standards for network operation, B5G and 6G integration and \ac{NTN}-specific protocols. This section outlines a standardization roadmap to ensure compatibility, scalability and interoperability of these methods in different \ac{NTN} applications.

\subsubsection{3GPP}

The 3rd Generation Partnership Project (3GPP) has laid the foundation for \ac{NTN} integration into the 5G ecosystem \cite{el2023introduction}. 3GPP Release 17 \& 18 has introduced \ac{NTN} support for 5G New Radio (NR) and IoT applications over satellite. Key areas for standardization has been as follows: \textit{(i)} Ensuring seamless handovers between terrestrial and non-terrestrial networks while maintaining consistent Quality of Service (QoS). \textit{(ii)} Adapting the 5G radio interface for satellite links, addressing issues such as latency and Doppler shift. \textit{(iii)} Standardizing the use of gNBs, \ac{NTN}-gNBs, and sNBs with regenerative and transparent payloads for satellite communication. At the same time, the design of probabilistic forecasts paradigm can be brought into line with these standards to enable efficient resource management and ensure compliance with \glspl{SLA}.

\subsubsection{ETSI}

The European Telecommunications Standards Institute (ETSI) defines Multi-Access Edge Computing (MEC) frameworks that can be adapted for \ac{NTN} environments \cite{kafle2024integrated}. Standardization efforts that could be focused on within the context of probabilistic forecasting are: \textit{(i)} Deploying predictive models at edge nodes, such as onboard satellites ground segment gateways or base stations, for real-time decision-making. \textit{(ii)} Integrating probabilistic forecasting models into MEC platforms for monitoring and enforcing \ac{SLA} compliance in \ac{NTN} deployments.

\subsubsection{ITU}

The International Telecommunication Union Radiocommunication Sector (ITU-R) regulates spectrum usage for \ac{NTN}s \cite{molleryd2025regulatory}. In this context, integrating probabilistic forecasting requires: \textit{(i) Dynamic Spectrum Allocation Standards:} Aligning forecasting models with spectrum-sharing protocols to predict and allocate frequencies dynamically based on demand. \textit{(ii) Interference Management:} Ensuring that forecasting-driven resource allocation adheres to ITU-R interference mitigation standards.

\subsubsection{ISO}

\ac{NTN}s operate in environments where data security and interoperability are critical. While ISO does not have a dedicated \ac{NTN}-specific standard, it works alongside ITU, 3GPP, and ETSI.  For this reason, ISO has close works related to telecommunications, including satellite and space networks. Key standardization areas in line with probabilistic forecasting in this domain can include: (i) Introduction of ISO-compliant data formats for telemetry, prediction models and network statistics in all \ac{NTN} systems. (ii) Integration of ISO/IEC 27001-compliant mechanisms \cite{ISO27001} into forecasting frameworks to protect sensitive data transmitted via integrated TNs-\ac{NTN}s.

\subsubsection{Cross-Sector Coordination}

Integrated \ac{TN}-\ac{NTN}s can also used in various industries, including maritime, aviation and IoT applications. In these carious domains, standardization can also consider the following aspects: (i) Harmonization of forecasting models with standards for communication in maritime transport (e.g. IMO), aviation (e.g. ICAO) and IoT (e.g. LoRaWAN) while relying on Integration of \ac{NTN} and TN (e.g. O-RAN). (ii) Development of cross-sector guidelines to ensure seamless integration of \ac{NTN} services with terrestrial, cloud and application-specific networks.

\subsection{Challenges, Mitigation Strategies and Future Directions}

The integration of TN (e.g. O-RAN) into \ac{NTN}s can  introduce unique challenges. First, \ac{NTN} links face high latency and Doppler shifts, which complicate resource provisioning. Predictive algorithms and AI-based compensation mechanisms are deployed to address these challenges. Second, the movement of satellites and UAS leads to rapidly changing network topologies.  However, real-time telemetry data and probabilistic forecasting can enable dynamic topology mapping. Third,  \ac{NTN} links often operate under strict bandwidth limitations. However, advanced frequency reuse and resource-sharing algorithms enhanced with probabilistic forecasting can ensure efficient bandwidth utilization. \textcolor{black}{Finally, AI-driven \ac{NTN} operations may face latency, frequency of model updates, and dynamic network conditions, requiring a trade-off between computational cost and accuracy to determine optimal AI/ML models based on available \ac{NTN} resources.} 

In future, the roadmap for levering probabilistic forecasting in integrated \ac{TN}-\ac{NTN} must also anticipate the following: (i) Preparing forecasting models for tighter \ac{NTN}-terrestrial integration in B5G and 6G networks, including the use of AI/ML for automated network management. (ii) Standardizing forecasting-driven allocation for high-frequency bands (e.g., Ka-band, mmWave, THz) to enable high-capacity \ac{NTN} applications.  (iii) Supporting the development of predictive resource-sharing frameworks for \glspl{ISL} within LEO constellations. Finally, standardized probabilistic forecasting ensures reliable, scalable TN-NTN services that meet diverse application needs.

\section{Conclusions}
\label{conclusions}

This paper presented a comprehensive framework for integrating \glspl{NTN} into the O-RAN architecture, leveraging probabilistic forecasting for efficient resource provisioning across terrestrial and non-terrestrial segments. We explored the general architecture and proposed a unified workflow to enable seamless coordination between TN and \ac{NTN} entities. Experiments show probabilistic forecasting outperforms deterministic methods in TN-NTN traffic prediction and provisioning. Additionally, we discussed the significance of coordinated resource allocation and cross-segment \ac{SLA} optimization to enhance overall network efficiency.  Finally, we aligned our model with standards while addressing TN-NTN challenges like latency, topology, and bandwidth constraints.

\section{Acknowledgment}
This work is partially supported by UNITY-6G project, funded from European Union’s Horizon Europe Smart Networks and Services Joint Undertaking (SNS JU) research and innovation programme under the Grant Agreement No 101192650; by the European Commission under the “5GSTARDUST” Project, which received funding from the SNS JU under the European Union’s Horizon Europe research and innovation programme under Grant Agreement No. 101096573,  by the SONATA project, funded by the grant CHIST-ERA-20-SICT-004 by PCI2021-122043-2A/AEI/10.13039/501100011033. 

\ifCLASSOPTIONcaptionsoff
  \newpage
\fi

\bibliographystyle{IEEEtran}
\bibliography{refs}

\begin{thebibliography}{10}
\providecommand{\url}[1]{#1}
\csname url@samestyle\endcsname
\providecommand{\newblock}{\relax}
\providecommand{\bibinfo}[2]{#2}
\providecommand{\BIBentrySTDinterwordspacing}{\spaceskip=0pt\relax}
\providecommand{\BIBentryALTinterwordstretchfactor}{4}
\providecommand{\BIBentryALTinterwordspacing}{\spaceskip=\fontdimen2\font plus
\BIBentryALTinterwordstretchfactor\fontdimen3\font minus \fontdimen4\font\relax}
\providecommand{\BIBforeignlanguage}[2]{{%
\expandafter\ifx\csname l@#1\endcsname\relax
\typeout{** WARNING: IEEEtran.bst: No hyphenation pattern has been}%
\typeout{** loaded for the language `#1'. Using the pattern for}%
\typeout{** the default language instead.}%
\else
\language=\csname l@#1\endcsname
\fi
#2}}
\providecommand{\BIBdecl}{\relax}
\BIBdecl

\bibitem{guidotti2024role}
A.~Guidotti, A.~Vanelli-Coralli, M.~El~Jaafari, N.~Chuberre, J.~Puttonen, V.~Schena, G.~Rinelli, and S.~Cioni, ``Role and evolution of non-terrestrial networks towards 6g systems,'' \emph{IEEE Access}, 2024.

\bibitem{hosseinian2021review}
M.~Hosseinian, J.~P. Choi, S.-H. Chang, and J.~Lee, ``Review of 5g ntn standards development and technical challenges for satellite integration with the 5g network,'' \emph{IEEE Aerospace and Electronic Systems Magazine}, vol.~36, no.~8, pp. 22--31, 2021.

\bibitem{polese2023understanding}
M.~Polese \emph{et~al.}, ``Understanding o-ran: Architecture, interfaces, algorithms, security, and research challenges,'' \emph{IEEE Communications Surveys \& Tutorials}, 2023.

\bibitem{el2023introduction}
M.~El~Jaafari, N.~Chuberre, S.~Anjuere, and L.~Combelles, ``Introduction to the 3gpp-defined ntn standard: A comprehensive view on the 3gpp work on ntn,'' \emph{International Journal of Satellite Communications and Networking}, vol.~41, no.~3, pp. 220--238, 2023.

\bibitem{campana2023ran}
R.~Campana, C.~Amatetti, and A.~Vanelli-Coralli, ``O-ran based non-terrestrial networks: Trends and challenges,'' in \emph{2023 Joint European Conference on Networks and Communications \& 6G Summit (EuCNC/6G Summit)}.\hskip 1em plus 0.5em minus 0.4em\relax IEEE, 2023, pp. 264--269.

\bibitem{Lar19}
L.~M.~P. Larsen, A.~Checko, and H.~L. Christiansen, ``{A Survey of the Functional Splits Proposed for 5G Mobile Crosshaul Networks},'' \emph{EEE Communications Surveys \& Tutorials}, vol.~21, no.~1, pp. 146--172, 2019.

\bibitem{Cam23}
R.~Campana, C.~Amatetti, and A.~Vanelli-Coralli, ``{O-RAN based Non-Terrestrial Networks: Trends and Challenges},'' in \emph{2023 Joint European Conference on Networks and Communications \& 6G Summit (EuCNC/6G Summit)}, 2023, pp. 264--269.

\bibitem{Dau24}
A.~Daurembekova and H.~D. Schotten, ``{Unified 3D Networks: Dynamic RAN Functions Placement and Link Challenges},'' in \emph{2024 International Symposium on Networks, Computers and Communications (ISNCC)}, 2024, pp. 1--6.

\bibitem{haupt2019use}
S.~E. Haupt, M.~G. Casado, M.~Davidson, J.~Dobschinski, P.~Du, M.~Lange, T.~Miller, C.~Mohrlen, A.~Motley, R.~Pestana \emph{et~al.}, ``The use of probabilistic forecasts: Applying them in theory and practice,'' \emph{IEEE Power and Energy Magazine}, vol.~17, no.~6, pp. 46--57, 2019.

\bibitem{10329915}
\textcolor{black}{Hoffmann, Marcin and others}, ``\textcolor{black}{Open RAN xApps Design and Evaluation: Lessons Learnt and Identified Challenges},'' \emph{\textcolor{black}{IEEE Journal on Selected Areas in Communications}}, vol. \textcolor{black}{42}, no. \textcolor{black}{2}, pp. \textcolor{black}{473--486}, \textcolor{black}{2024}.

\bibitem{kasuluru2023use}
V.~Kasuluru, L.~Blanco, and E.~Zeydan, ``On the use of probabilistic forecasting for network analysis in open ran,'' in \emph{2023 IEEE International Mediterranean Conference on Communications and Networking (MeditCom)}.\hskip 1em plus 0.5em minus 0.4em\relax IEEE, 2023, pp. 258--263.

\bibitem{kasuluru2024impact}
V.~Kasuluru, L.~Blanco, C.~J. Vaca-Rubio, and E.~Zeydan, ``On the impact of prb load uncertainty forecasting for sustainable open ran,'' in \emph{2024 IEEE 35th International Symposium on Personal, Indoor and Mobile Radio Communications (PIMRC)}.\hskip 1em plus 0.5em minus 0.4em\relax IEEE, 2024, pp. 1--7.

\bibitem{kafle2024integrated}
V.~P. Kafle, M.~Sekiguchi, H.~Asaeda, and H.~Harai, ``Integrated network control architecture for terrestrial and non-terrestrial network convergence,'' \emph{IEEE Communications Standards Magazine}, vol.~8, no.~1, pp. 12--19, 2024.

\bibitem{molleryd2025regulatory}
B.~M{\"o}lleryd, M.~Ozger, M.~Westring, A.~Nordl{\"o}w, D.~Schupke, U.~Engstr{\"o}m, C.~Cavdar, M.~Lindborg, and N.~Sciammetta, ``Regulatory and spectrum policy challenges for combined airspace and non-terrestrial networks,'' \emph{Telecommunications Policy}, vol.~49, no.~1, p. 102875, 2025.

\bibitem{ISO27001}
\BIBentryALTinterwordspacing
ISO, \emph{{ISO/IEC 27001:2022} Information security, cybersecurity and privacy protection — Information security management systems — Requirements}, International Organization for Standardization Std., 2022. [Online]. Available: \url{https://www.iso.org/standard/27001}
\BIBentrySTDinterwordspacing

\end{thebibliography}
\end{document}